\documentclass[10pt,aps,prl,twocolumn,superscriptaddress]{revtex4}
\usepackage[linktocpage=true, colorlinks=true, urlcolor=blue, linkcolor=blue, citecolor=blue]{hyperref}
\usepackage{graphicx}
\usepackage{amsmath,amssymb}
\usepackage{color}
\usepackage{url}
\usepackage{hyperref}
\usepackage{cleveref}
\usepackage{xfrac}
\usepackage{upgreek}
\usepackage{epstopdf}

\graphicspath{{./Images/}}

\begin{document}

\title{Reverse contrast laser Doppler holography for lower frame rate retinal and choroidal blood flow imaging}

\author{L\'eo Puyo}
\affiliation{Corresponding author: gl.puyo@gmail.com}
\affiliation{Centre Hospitalier National d'Ophtalmologie des Quinze-Vingts, INSERM-DHOS CIC 1423. 28 rue de Charenton, 75012 Paris France}
\affiliation{Institut de la Vision-Sorbonne Universit\'es. 17 rue Moreau, 75012 Paris France}

\author{Michel Paques}
\affiliation{Centre Hospitalier National d'Ophtalmologie des Quinze-Vingts, INSERM-DHOS CIC 1423. 28 rue de Charenton, 75012 Paris France}
\affiliation{Institut de la Vision-Sorbonne Universit\'es. 17 rue Moreau, 75012 Paris France}

\author{Michael Atlan}

\affiliation{Institut Langevin. Centre National de la Recherche Scientifique (CNRS). Paris Sciences \& Lettres (PSL University). \'Ecole Sup\'erieure de Physique et de Chimie Industrielles (ESPCI Paris) - 1 rue Jussieu. 75005 Paris France}

\date{\today}

\begin{abstract}
Laser Doppler holography (LDH) is an interferometric blood flow imaging technique based on full-field measurements of the Doppler spectrum. LDH has so far been demonstrated in the retina with ultrafast cameras, typically at 75 kHz. We show here that a similar method can be implemented with camera frame rates 10 times slower than before. Thanks to energy conservation, low and high frequency local power Doppler signals have opposite variations, and a simple contrast inversion of the low frequency power Doppler reveals fast blood flow beyond the camera detection bandwidth for conventional laser Doppler measurements. Relevant blood flow variations and color composite power Doppler images can be obtained with camera frame rates down to a few kHz.
\end{abstract}

\maketitle

Laser Doppler holography (LDH) is a digital holographic method where blood flow is measured from the interference of coherent light backscattered by the eye with a reference beam~\cite{SimonuttiPaquesSahel2010, MagnainCastelBoucneau2014, Pellizzari2016}. The Doppler broadening can be measured over the full-field array of an ultrafast camera thanks to the coherent gain brought by the reference arm. The power Doppler is calculated pixel-wise as the integral of the high-pass filtered Doppler power spectrum density (DPSD) to reveal blood flow from the larger Doppler broadening of light scattered in blood vessels. By using a sliding short-time window, variations of retinal blood flow over cardiac cycles can be measured in power Doppler units with a few milliseconds of temporal resolution~\cite{Puyo2018}. The method was previously introduced in ophthalmology to image retinal and choroidal blood flow in the human eye~\cite{Puyo2018, Puyo2019, Puyo2019b}. LDH measurements demonstrated so far have been performed with ultrahigh camera frame rates, as required to sample Doppler shifts up to a few tens of kHz. Undersampling the Doppler broadening otherwise leads to erroneous blood flow measurements in large vessels lumen or during systole~\cite{Puyo2018}. This frame rate requirement is a serious limitation for the technique as ultrafast camera are expensive and challenging to use. In this Letter, we demonstrate that qualitatively similar blood flow images can be obtained with frame rates 10 times slower than before by taking advantage of the energy conservation of the local DPSD. We have found that low and high frequency local power Doppler signals have opposite variations, and that simply reversing the contrast of the very low frequency power Doppler allows to obtain relevant blood flow measurements with camera frame rates down to a few kHz. This scheme allows to reveal fast blood flow that is largely beyond the camera detection bandwidth for conventional laser Doppler measurements~\cite{Tanaka1974, Stern1975}. The method is first validated by comparing the low and high frequency content of ultrahigh speed LDH measurements, and then it is demonstrated with real low frame rate acquisitions.

We use the LDH setup presented in~\cite{Puyo2018} to image blood flow in the posterior pole of human eyes. Informed consent was obtained from the subjects, experimental procedures adhered to the tenets of the Declaration of Helsinki, the study authorization was obtained from the appropriate local ethics review boards CPP and ANSM, and the clinical trial was registered under the references IDRCB 2019-A00942-5, and NCT04129021. Interferograms are recorded using a CMOS camera (Ametek - Phantom V2511) used in a 512 $\times$ 512 format with ultrafast (i.e. between 60 and 75 kHz), and slow (i.e. between 4 and 10 kHz) frame rates. The ocular exposure to the 785 nm single frequency laser diode is constant, and the camera exposure time is set to the maximum possible according to its sampling frequency. The data processing is essentially the same for both the fast and slow frame rates. The digital holograms are numerically reconstructed by angular spectrum propagation, and analyzed by short-time Fourier transform~\cite{Puyo2018}. In order to remove the Doppler contribution of eye motion, each short-time window is first filtered by singular value decomposition (SVD), as detailed elsewhere~\cite{Puyo2020}. The eigenvectors of the holograms space-time matrix associated to the eigenvalues of highest energy are rejected with an equivalent Fourier cutoff of 0.2 kHz, which allows to access low frequency blood flow signals, otherwise dominated by spurious contributions. Then, for each time point $t_{n}$ the power Doppler $M_{0}$ is calculated as the integral over a given frequency range of the power spectrum density $S$ calculated as the squared magnitude of the holograms temporal Fourier transform: $M_{0}(x,y,t_{n}) = \int S(x,y,t_{n},f) \;  df$. Finally, power Doppler images are corrected for non-uniform illumination, and the baseline signal is subtracted from the power Doppler movie~\cite{Puyo2019b}. In order to study blood flow variations with low frame rate acquisitions, the negative power Doppler is considered, which amounts to reversing the contrast of images (inversion of the image grayscale).
\begin{figure}[t!]
\centering
\includegraphics[width = 1\linewidth]{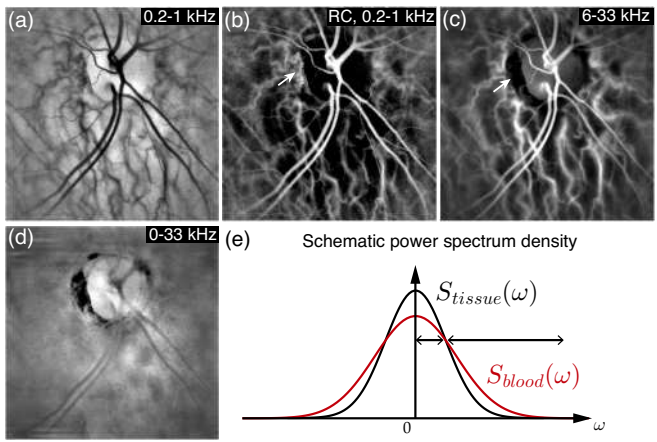}
\caption{$f_S = 67 \, \rm kHz$ LDH measurement. (a) 0.2-1 kHz power Doppler image. (b) 0.2-1 kHz RC power Doppler. (c) The 6-33 kHz power Doppler exhibits very similar contrast. (d) 0-33 kHz power Doppler. (e) Energy conservation in LDH: a broader spectrum leads to an energy depletion in the low frequency range, so low and high frequency power Doppler images are oppositely contrasted. \textcolor{blue}{\href{https://youtu.be/2za0KnbSmS8}{Visualization 1}}.
}
\label{fig_1_HyperspectralSchematic}
\end{figure}
This process is illustrated in Fig.~\ref{fig_1_HyperspectralSchematic} with a LDH measurement performed at $f_S = 67 \; \rm kHz$. As noted in previous work~\cite{Puyo2018}, blood vessels are visible with a dark contrast on low frequency power Doppler images. This is shown in Fig.~\ref{fig_1_HyperspectralSchematic}(a) with a 0.2-1 kHz power Doppler image: the power Doppler is lower in vascularized structures than in the surrounding tissue. In Fig.~\ref{fig_1_HyperspectralSchematic}(b) is shown the reversed contrast (RC) of the same image, and the vasculature is now contrasted almost exactly like on the standard high frequency (6-33 kHz) power Doppler image shown in Fig.~\ref{fig_1_HyperspectralSchematic}(c). Choroidal vessels can be observed on both images, whereas they cannot be observed when integrating the entire DPSD in Fig.~\ref{fig_1_HyperspectralSchematic}(d), showing that the higher absorption in blood vessels is a negligible source of contrast for the vasculature in Fig.~\ref{fig_1_HyperspectralSchematic}(a). Instead this phenomenon is better explained by energy conservation. All pixels correspond to fundus areas that have received the same amount of energy from the laser beam illumination, and the Parseval-Plancherel identity states that the total power Doppler (area under the curve of the power spectrum density) is a quantity that is conserved in the temporal and spectral domains. Thus, aside from the local difference of reflectivity and absorption, the total area under the curve of the DPSD should be the same everywhere because it corresponds to the laser irradiance. A faster flow in blood vessels leads to a broader DPSD because of greater Doppler shifts induced by faster scattering particles.  The schematic in Fig.~\ref{fig_1_HyperspectralSchematic}(e) illustrates the spectral energy depletion at low frequency in areas with faster flows. Conversely, there is less Doppler broadening occurring in areas with slower flows, especially in tissue, so the power spectrum density remains higher at low frequency. Therefore, because the power spectrum density at high frequency is greater in fast flow areas than in tissue, it is also greater in tissue than in blood flow areas at low frequency. A larger Doppler broadening, which is visible with a positive contrast on high frequency power Doppler images, is also visible on low frequency power Doppler images with a negative contrast. In other words, low and high frequency power Doppler images are oppositely contrasted. The differences of reflectivity and absorption of the various structures also come into play and invalidate an exact energy conservation. For example the low frequency RC power Doppler image does not reflect the content of the high frequency power Doppler image in the area pointed by the arrow in Fig.~\ref{fig_1_HyperspectralSchematic}(b) and (c): the absorption is locally greater because of the exposed retinal pigment epithelium (RPE). The strong absorption in the RPE negates the energy conservation, which is why it is not observed with the same contrast. However, the experimental results shown in this Fig. demonstrate that as far as blood flow images are concerned, the low frequency RC power Doppler can be exploited to satisfactorily reveal the high frequency power Doppler content. In \textcolor{blue}{\href{https://youtu.be/2za0KnbSmS8}{Visualization 1}}, we show that faster flows are revealed by higher frequencies both in standard power Doppler over the 1 to 33 kHz range, and in RC power Doppler over the 0.2 to 4 kHz range. Interestingly the two signals are mixed-up in the 1-4 kHz range. This kind of pattern was systematically observed in more than 1000 measurements performed in close to 300 subjects.

\begin{figure}[t!]
\centering
\includegraphics[width = 1\linewidth]{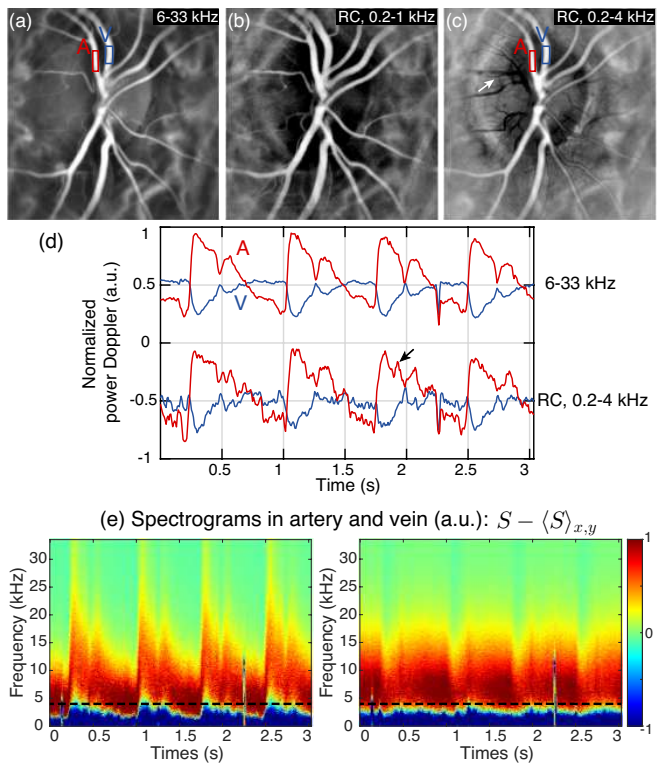}
\caption{$f_S = 67 \, \rm kHz$, measuring pulsatile blood flow with RC low frequency power Doppler. (a) 6-33 kHz power Doppler image. (b) 0.2-1 kHz RC power Doppler. (c) 0.2-4 kHz RC power Doppler. (d) Variations in an artery and a vein. Movies in \textcolor{blue}{\href{https://youtu.be/HFphDhQ13tY}{Visualization 2}}. (e) Spectrograms in the same areas.
}
\label{fig_2_HighFramerateProofOfConcept}
\end{figure}

In Fig.~\ref{fig_2_HighFramerateProofOfConcept}, we demonstrate that the RC low frequency power Doppler also reflects dynamically the content of the high frequency power Doppler and reveals variations induced by pulsatile blood flow. With a LDH acquisition performed at $f_S = 67 \, \rm kHz$, we have measured in a retinal artery and vein the variations of RC low frequency power Doppler, and of standard high frequency power Doppler. The 0.2-1 kHz range is enough to produce satisfying perfusion maps, but because pulsatile flow is faster than the steady flow, pulsatile variations must be monitored with a higher frequency range. Empirically we have determined that the 0.2-4 kHz frequency range is sufficient to mirror in RC the blood flow variations measured with the 6-33 kHz range, so we selected large retinal vessels whose blood flow is adequately revealed with the 6-33 kHz range. Each pair of artery/vein signals was normalized and centered based on the standard deviation and average values of the arterial signal. The corresponding power Doppler images in Fig.~\ref{fig_2_HighFramerateProofOfConcept}(a) and (c) show the selected areas of interest. Some vessels appearing with a dark contrast in Fig.~\ref{fig_2_HighFramerateProofOfConcept}(c) are not visible with either the 6-33 kHz image, or with the 0.2-1 kHz RC power Doppler in Fig.~\ref{fig_2_HighFramerateProofOfConcept}(b). These dark vessels are revealed because they carry slow flow revealed by positive power Doppler over the 1-4 kHz frequency range. As shown in Fig.~\ref{fig_2_HighFramerateProofOfConcept}(d), the blood flow variations measured with both methods are strikingly similar in both vessels, and even the relative amplitudes of the artery/vein variations are preserved. The variations are slightly noisier with the low frequency band, possibly because it is averaged over a smaller ensemble length in Fourier space, and some minor perturbation probably resulting from residual bulk motion not totally suppressed by the SVD filtering can also be observed (arrow). In Fig.~\ref{fig_2_HighFramerateProofOfConcept}(e) are shown the spectrograms measured in the same areas, from which the spatially averaged spectrum was subtracted. The variations of power Doppler below the 4 kHz cutoff (horizontal dashed line) negatively follow the pulsatile variations of the spectrum at high frequency. Over the course of cardiac cycles, as blood flow is accelerated, the energy is shifted towards higher frequencies and the low frequency power Doppler is depleted.

\begin{figure}[t!]
\centering
\includegraphics[width = 1\linewidth]{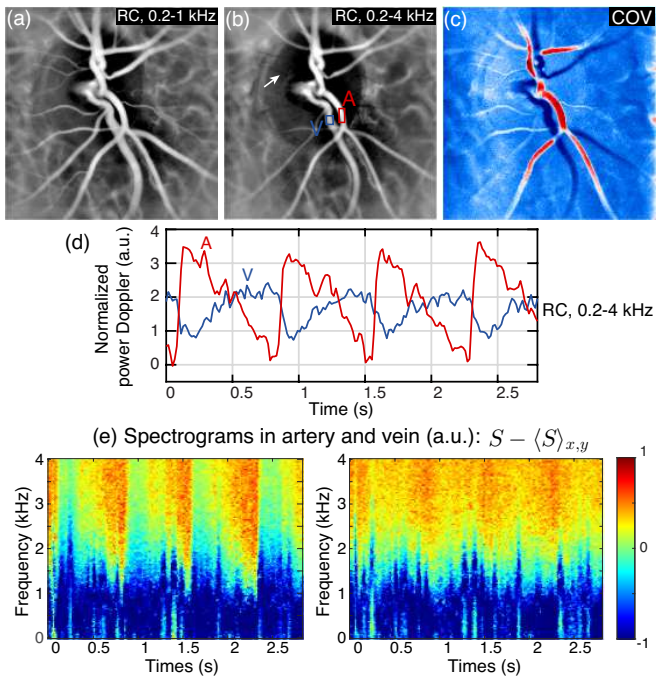}
\caption{$f_S = 8 \, \rm kHz$ LDH measurement. (a) 0.2-1 kHz RC power Doppler image. (b) 0.2-4 kHz RC power Doppler. (c) The local power Doppler coefficient of variation differentiates arteries and veins. (d) The RC power Doppler variations in arteries and veins exhibit the waveforms typically observed with high frequency LDH measurements. Movie in \textcolor{blue}{\href{https://youtu.be/hOyqRTL7Tfg}{Visualization 3}}. (e) Spectrograms in the same areas.
}
\label{fig_3_LowFrequencyMeasurePLOTs}
\end{figure}

The results shown in Fig.~\ref{fig_1_HyperspectralSchematic} and ~\ref{fig_2_HighFramerateProofOfConcept} suggest that using a low frame rate and reversing the power Doppler contrast could be used to perform blood flow measurements similar to those obtained from ultrafast frame rates LDH acquisitions. This technique is explored in Fig.~\ref{fig_3_LowFrequencyMeasurePLOTs} with a LDH measurement recorded at low frame rate. The sampling frequency was set to $f_S = 8 \, \rm kHz$ to access the 0.2-4 kHz frequency band. It is possible to see on the 0.2-1 kHz RC power Doppler image in Fig.~\ref{fig_3_LowFrequencyMeasurePLOTs}(a) that the perfusion map of fast blood flow can indeed be revealed as expected. Similarly as before, the 0.2-4 kHz RC power Doppler image in Fig.~\ref{fig_3_LowFrequencyMeasurePLOTs}(b) shows the perfusion map with some small vessels appearing with a dark contrast (arrow). The corresponding variations of negative power Doppler in an artery (red) and a vein (blue) are plotted in Fig.~\ref{fig_3_LowFrequencyMeasurePLOTs}(d). The standard systolodiastolic flow variations and the dicrotic notch can be observed in the retinal artery, as well as a cycloidal flow profile in the vein similar to what is typically measured with high frequency LDH. The coefficient of variation map that was previously introduced to differentiate retinal arteries and veins is shown in Fig.~\ref{fig_3_LowFrequencyMeasurePLOTs}(c), and exhibits a quality similar to what was previously obtained at high frequency~\cite{Puyo2019b}. Finally, in Fig.~\ref{fig_3_LowFrequencyMeasurePLOTs}(e) are shown the spectrograms measured in the same areas, from which the spatially averaged spectrum was subtracted. Only a small portion of the Doppler spectrum is measured, but the pulsatile shift of energy on which RC measurements rely is visible. The results presented in this Fig. demonstrate that perfusion maps and blood flow variations of fast blood flow largely beyond the detection bandwidth of the camera can be revealed by inverting the contrast of low frame rate LDH measurements. The images shown in Fig.~\ref{fig_1_HyperspectralSchematic} and \ref{fig_2_HighFramerateProofOfConcept} are obtained from high frequency LDH measurements, where the Doppler broadening is correctly sampled and the measured power spectrum density reflects the real Doppler spectrum. When the measurement is carried out at low frequency, the undersampled part of the spectrum is aliased, and it could be expected that the ability of the RC low frequency power Doppler to reflect positive high frequency power Doppler would be jeopardized. However the results presented in Fig.~\ref{fig_3_LowFrequencyMeasurePLOTs} obtained with low frame rate RC-LDH indicate that it seems not to be the case. The undersampled part of the Doppler spectrum may be sufficiently attenuated by the sampling sinc frequency response.

\begin{figure}[t!]
\centering
\includegraphics[width = 1\linewidth]{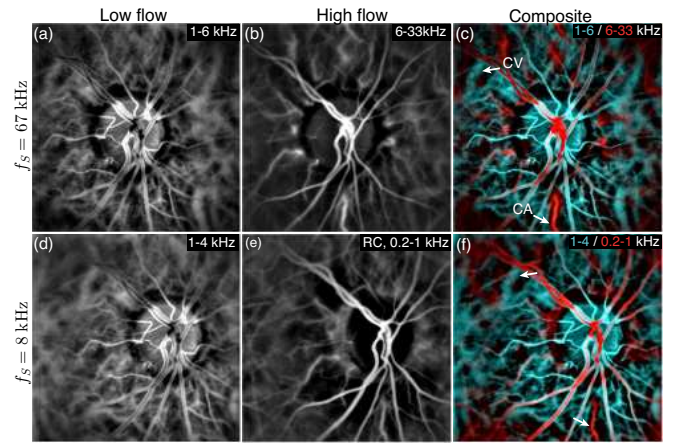}
\caption{Making low/high flow composite images with high and low frame rates. (a), (b), and (c) show the usual composite images previously demonstrated~\cite{Puyo2019}. (d), (e), and (f) show the result with a 8 kHz frame rate: the high flow image is obtained from the RC low frequency power Doppler. The composite movies are shown in \textcolor{blue}{\href{https://youtu.be/VCUJbFDZr3Q}{Visualization 4}} and \textcolor{blue}{\href{https://youtu.be/I4kSQBgbKAQ}{Visualization 5}}.
}
\label{fig_4_LowFrequencyMeasureCOMPOSITES}
\end{figure}

In our previous work~\cite{Puyo2019, Puyo2019b}, we showed that composite power Doppler images from low and high frequency ranges that reveal slow and fast blood on a single image can be very useful. The typical frequency ranges used for these images are 1-6 kHz and 6 kHz - $f_S / 2$, where $f_S$ is usually between 60 and 75 kHz. Arterial and venous choroidal vessels are differently contrasted on these composite images, thanks to the natural discrepancy of flow between these vessels~\cite{Puyo2019}. We demonstrate in Fig.~\ref{fig_4_LowFrequencyMeasureCOMPOSITES} that composite images with very similar contrast can be generated from RC low frequency LDH measurement. LDH measurements were performed in the optic nerve head (ONH) of a same eye with sampling frequencies of 67 kHz and 8 kHz. In Fig.~\ref{fig_4_LowFrequencyMeasureCOMPOSITES}(a) and (b) are shown the usual power Doppler images for the low and high flow for $f_S = 67 \; \rm \ kHz$, and in Fig.~\ref{fig_4_LowFrequencyMeasureCOMPOSITES}(c) the resulting composite. Then for the measurement at $f_S = 8 \; \rm \ kHz$, we show in Fig.~\ref{fig_4_LowFrequencyMeasureCOMPOSITES}(d) that a similar low flow power Doppler image is obtained from the 1-4 kHz frequency range. Then, as shown in Fig.~\ref{fig_4_LowFrequencyMeasureCOMPOSITES}(e), the 0.2-1 kHz RC power Doppler image exhibits the same contrast as the very high frequency power Doppler, which reveals the fast flow. Consequently, combining these two images into the color composite Doppler image displayed in Fig.~\ref{fig_4_LowFrequencyMeasureCOMPOSITES}(f) leads to an image contrasted very similarly to the composite image obtained with the 67 kHz measurement. Retinal vessels can be observed on both images with the same contrast. More importantly, as shown by the arrows 'CA' and 'CV', the choroidal artery and vein in the vicinity of the ONH are similarly differentiated on both composite images. Thus, colored composite images of equivalent quality to those obtained from ultrafast LDH measurements can be obtained from low frame rate LDH measurements. In the associated Visualizations however, some perturbation (flashes) are observed with the low frame rate measurement, again probably from residual eye motion not totally suppressed by the SVD.

\begin{figure}[]
\centering
\includegraphics[width = 1\linewidth]{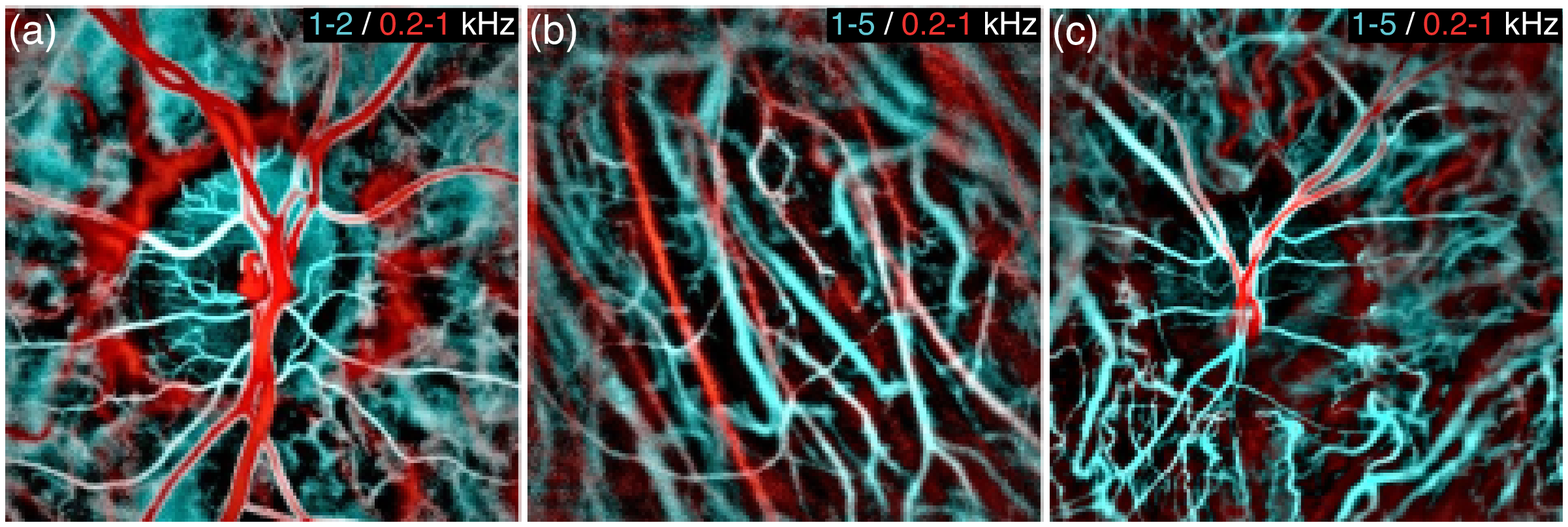}
\caption{Application of low frame rate RC-LDH. (a) $f_S = 4 \, \rm kHz$, ONH region. (a) $f_S = 10 \, \rm kHz$, peripheral region, \textcolor{blue}{\href{https://youtu.be/v---F7RvFfU}{Visualization 6}}. (c) $f_S = 10 \, \rm kHz$, case of vascular impairment, the reduced flow in the lower hemivessels is evidenced, \textcolor{blue}{\href{https://youtu.be/JfZ1fH1m6sA}{Visualization 7}}.
}
\label{fig_5_LowFrequencyComposites}
\end{figure}

In Fig.~\ref{fig_5_LowFrequencyComposites}, we show blood flow measurement with RC-LDH operating at low frame rate. The sampling frequency is first set to $f_S = 4 \, \rm kHz$ to image a healthy ONH in Fig.~\ref{fig_5_LowFrequencyComposites}(a). The Doppler color composite image has been obtained by fusing the 0.2-1 kHz RC power Doppler and the 1-2 kHz power to reveal fast (red) and low (cyan) blood flow. The resulting image reveals both the ONH microvasculature and fast blood flow in large vessels in the expected color tones. The image of a peripheral region in Fig.~\ref{fig_5_LowFrequencyComposites}(b) from a 10 kHz measurement is able to differentiate choroidal vessels of similar size based on their flow. Finally in Fig.~\ref{fig_5_LowFrequencyComposites}(c), a vascular impairment is imaged with 10 kHz RC-LDH. The subject eye has received radiation therapy to treat a tumor below the ONH. The composite image obtained by fusing the 0.2-1 kHz RC power Doppler in red and the 1-5 kHz power Doppler in cyan clearly evidences the decreased perfusion in the inferior retinal hemivessels.

The frequencies in the 1-4 kHz range can be used to reveal both fast and slow blood flow, respectively with RC or normal power Doppler. Thus it is advantageous to sample the frequency at 8-10 kHz for the dual purpose of measuring the pulsatile variations of blood flow with RC power Doppler, and revealing slow blood flow with normal power Doppler. This frequency range is very affected by spurious interferometric contributions, so the use of a spatio-temporal filtering is critical to reveal the signal. It is particularly important to filter the pulsatile axial motion of the eye because they are synchronous with blood flow and occur in the direction that maximizes the Doppler effect~\cite{Puyo2020}. A limitation of low frequency RC-LDH is that it cannot offer absolute spectral measurements as high frequency LDH, but considering the significant reduction of camera frame rate, the similarity between the obtained blood flow traces and color composite images at low and high speed LDH is satisfying. For measurements of a same duration, the quality of blood flow measurements seems slightly lower with RC power Doppler, possibly because there are less pixels involved in the signal detection. However this is counterbalanced by the possibility to perform longer acquisitions. The quality of the perfusion maps, blood flow variations, color composite Doppler images obtained by RC-LDH seems of sufficient quality for clinical use.

In conclusion, the detection of low frequency light fluctuations by digital holography enables an efficient characterization of Doppler spectrum content at high frequency, beyond the Shannon-Nyquist sampling limit. Relevant perfusion maps and pulsatile blood flow variations of flow velocities largely out of the reach of conventional wideband laser Doppler instruments can be measured with RC-LDH. Color composite Doppler images of low/high flow can also be produced with surprisingly low camera frame rates.

\section*{Funding Information}
European Research Council (Synergy HELMHOLTZ \#610110). Institut hospitalo-universitaire (FOReSIGHT, ANR-18-IAHU-01). Region Ile-de-France (Sesame program, 4DEye project). The Titan RTX used for this research was donated by the NVIDIA Corporation.

\section*{Disclosures}
The authors declare no conflicts of interest.


\section*{Supplementary Material}
\vspace{-2em}
\noindent

\begin{center}
\textcolor{blue}{\href{https://youtu.be/bC8TOqwxrQo}{Supplementary Visualization 1}} \\
\textcolor{blue}{\href{https://youtu.be/6EppAUNQiOU}{Supplementary Visualization 2}} \\
\textcolor{blue}{\href{https://youtu.be/gnpjaZEbLNY}{Supplementary Visualization 3}}

\end{center}

%

%

\end{document}